\documentstyle[prl,aps,twocolumn]{revtex}

\catcode`\@=11

\def\maketitle2{\par 
\begingroup
\let\cite\@bylinecite
\def\thefootnote{\fnsymbol{footnote}}%
\twocolumn[\@maketitle2\vskip2pc]%
\thispagestyle{plain}\@thanks
\endgroup
\def\thefootnote{\arabic{footnote}}%
\setcounter{footnote}{0}%
\let\maketitle2\relax \let\@maketitle2\relax
\let\@thanks\relax \let\@authoraddress\relax \let\@title\relax
\let\@date\relax \let\thanks\relax \let\@abstract\relax 
\let\@pacs\relax}

\def\abstract#1{\gdef\@abstract{{\par 
\bgroup
\ifdim\prevdepth=-1000pt \prevdepth0pt\fi
\hsize\columnwidth
\dimen0=-\prevdepth \advance\dimen0 by17.5pt \nointerlineskip
\small\vrule width 0pt height\dimen0 \relax}{~~}#1\egroup}}

\def\pacs#1{\gdef\@pacs{{\par 
\bgroup
\hsize\columnwidth \parindent0pt
\ifdim\prevdepth=-1000pt \prevdepth0pt\fi
\dimen0=-\prevdepth \advance\dimen0 by20pt\nointerlineskip
\egroup} PACS numbers:~#1}}

\def\@maketitle2{
\@preprint
\@title
\ifdim\prevdepth=-1000pt \prevdepth0pt\fi
\@authoraddress
\@date
\begin{list}{}{\leftmargin=0.10753\textwidth \rightmargin=\leftmargin
\itemsep=1pc\partopsep=-1pc}
\item\@abstract
\item\@pacs
\end{list}
}

\catcode`\@=12

\begin{document}

\draft

\preprint{LA-UR 96-3873}

\title{Cold, dilute, trapped bosons as an open quantum system}

\author{James Anglin\cite{email}}
\address{Theoretical Astrophysics, MS B288, Los Alamos National Laboratory,
Los Alamos, New Mexico 87545}

\abstract{
We present a master equation governing the reduced density
operator for a single trapped mode of a cold, dilute, weakly interacting
Bose gas; and we obtain an operator fluctuation-dissipation relation
in which the Ginzburg-Landau effective potential plays a physically
transparent role.  We also identify a decoherence effect that tends
to preserve symmetry, even when the effective potential has a ``Mexican
hat'' form.}

\pacs{03.75.Fi, 05.30.Jp, 11.30.Qc, 34.40.+n}

\maketitle2

\narrowtext

The recent observations of Bose condensation in magnetically trapped
gases\cite{t1,t2,t3,t4,t5} have brought phenomena hitherto reserved to
condensed matter physics into the interdisciplinary field now growing
between quantum optics and mesoscopic physics.  In particular, this
breakthrough may offer a weakly interacting system in which to study one
of the most interesting of quantum phenomena: the spontaneous breaking
of number eigenstates, which are invariant (up to a Hilbert space phase)
under the rotation $\hat{U} = e^{i\theta\hat{n}}$, into coherent states,
which transform as $\hat{U}|\alpha\rangle = |e^{i\theta}\alpha\rangle$.
Such symmetry breaking has long been thought to be a basic cause of
superfluidity and superconductivity.  Yet the breaking of the symmetry
whose Noether charge is particle number is still rather more mysterious
than the symmetry breaking involved in, say, the Standard Model of
particle theory, because instead of being due to an instability in the
microscopic Hamiltonian of the system, it is thought to be driven by an
{\it effective} potential of the form
\begin{equation}
V = (E_1-\mu)|\psi|^2 + E_2 |\psi|^4\;,
\end{equation}
where $\psi$ is a second-quantized destruction operator.

While it has long been customary to include an effective potential
of this form in the Hamiltonian for a gas of weakly interacting
bosons\cite{long}, the presence of the chemical potential $\mu$ must
remind us that this effective potential (EP) is fundamentally a property
of the system's thermal environment.  In non-equilibrium dynamics,
therefore, the EP should most properly appear, like temperature,
in the fluctuation-dissipation relation (FDR).  To realize such a
result, we must consider Bose condensation, and the possibly concomitant
symmetry breaking, to occur in an open quantum system.  The open system
paradigm is not only natural for gases that are cooled by evaporation and
thereafter steadily leak out of their traps; it is in general a powerful
{\em lingua franca} for describing the rapidly growing common ground
between quantum optics, mesoscopic quantum mechanics, and condensed
matter physics.  Recent work using this paradigm has developed analogs
of laser theory for atoms\cite{Guzmanetal,WisemanCollett,Hollandetal}
and excitons\cite{ImamogluRam}.  The nonlinearity of the interaction
between hard spheres makes open system calculations difficult for gases
in general, but in this Letter we present an idealized model of a cold,
dilute, trapped Bose gas, in which the problem may be solved analytically
to leading order in small parameters.  We are thereby able to
obtain the fully nonlinear, fully quantum mechanical FDR, and so make
contact between the open system approach and traditional many-body
formulations based on an EP.

Instead of the quadratic confining potentials of real experiments,
we imagine a Bose gas subjected to a deep but narrow spherical square
potential well, tuned so as to possess exactly one single-particle bound
state.  In the second-quantized formalism, this single-particle state
becomes a normal mode, with a discrete set of energy levels.  We treat
this mode as an open quantum system, interacting, via a short-ranged
two-particle potential, with a reservoir consisting of the continuum of
unbound modes.  We then outline a derivation (to be presented in detail
elsewhere) that uses Feynman's ordered operator calculus\cite{Feynman} to
obtain a Markovian master equation for the reduced density operator of
the bound mode.  This equation contains condensate growth and depletion
terms whose relative strength is characterized by an {\em operator}
fluctuation-dissipation relation, which is physically transparent,
and in which the Ginzburg-Landau EP may be discerned.  We also find an
additional phase diffusion term, representing quantum decoherence due
to scattering of unbound particles by the condensate.  We will discuss
the implications of this term for symmetry breaking at the end of this
Letter.

We therefore begin with the second-quantized Hamiltonian
\begin{eqnarray}\label{Hx}
\hat{H} &=& {\hbar^2\over2m}\!\!\int\!dV[
\vec\nabla\hat{\psi}^\dagger \cdot\vec\nabla\hat{\psi}
- 2\lambda^2\theta(\pi-\lambda r)\hat{\psi}^\dagger\hat{\psi}\nonumber\\
&&\qquad\qquad\qquad 
+ 4\pi a \hat{\psi}^\dagger\hat{\psi}^\dagger\hat{\psi}\hat{\psi}]\;,
\end{eqnarray}
where $\theta$ is the step function, and $a$ is the
scattering length of the gas particles ($\sigma = 16\pi a^2$ being the
cross section for two hard-sphere bosons scattering with momentum
transfer small compared to $a^{-1}$).
We then diagonalize the quadratic part of $\hat{H}$, by defining
$\hat{\psi}(\vec{r}) = \sum_{klm} \hat{\psi}_{klm} u_{klm}(\vec{r})$,
such that
\begin{equation}\label{modes} 
(\nabla^2 + k^2) u_{klm} = 2\lambda^2
\theta(\pi - \lambda r) u_{klm}\;.  
\end{equation} 
The $u_{klm}$ with real $k$ are unbound modes, scattered by the
well\cite{CT}.  We will let the spectrum of $k$ approach a continuum
by implicitly assuming a boundary condition at large radius.  There is
also exactly one bound solution, which is spherically symmetric; by
solving a transcendental equation we can compute its binding energy to
be $E_b \doteq 0.457\ {\hbar^2\lambda^2\over2m}$.  We denote this bound
mode wave function by $u_B(\vec{r})$, and its creation and destruction
operators by $\hat{\psi}_B^\dagger$ and $\hat{\psi}_B$.  The bound
mode will then constitute the observed system, and the continuum modes
the environmental reservoir, which will be in self-equilibrium with
temperature $(k_B\beta)^{-1}$ and chemical potential $\mu$, though it
may be far from equilibrium with the bound mode.

We then re-write the Hamiltonian (\ref{Hx}) in terms of the normal
modes, splitting it into bound mode, continuum, and interaction parts:
\begin{eqnarray}
\hat{H} &=& \hat{H}_B + \hat{H}_C + \hat{H}_I\nonumber\\
\hat{H}_B &=& -E_b\hat{n}_B + E_r \hat{n}_B(\hat{n}_B-1)\nonumber\\
\hat{H}_C &=& {\hbar^2\over2m}\sum_{klm}
	k^2\hat{\psi}^\dagger_{klm}\hat{\psi}_{klm} + \hat{H}_{scat}
	\equiv \hat{H}_{kin} + \hat{H}_{scat}\;,
\end{eqnarray}
where $\hat{n}_B \equiv \hat{\psi}^\dagger_B \hat{\psi}_B$, and the
condensate self-repulsion energy $E_r$ is of order $a\hbar^2\lambda^3
m^{-1}$, proportional to $\int\!d^3r\,u_B^4$.  $\hat{H}_{scat}$ is
a quartic operator which produces two-particle scattering among the
continuum modes.  It will be convenient to use the interaction picture,
in which operators evolve under $\hat{H}_B$ and $\hat{H}_C$:
\begin{eqnarray}\label{intpic}
\hat{\psi}_B(t) &=& e^{{i\over\hbar}(E_b-2E_r\hat{n}_B)t}\hat{\psi}_B(0) 
\equiv e^{-i\hat{\Delta} t} \hat{\psi}_B(0)\nonumber\\
\hat{\psi}_{klm}(t) &=& e^{-i{\hbar k^2\over2m}t} \hat{\Psi}_{klm}(t)\;.
\end{eqnarray}
Note that $\hat{\Delta}=\Delta(\hat{n})$ is simply the frequency difference
between the $n$th and the $(n-1)$th eigenstate of $\hat{H}_B$.
The operators $\hat{\Psi}_{klm}(t)$ evolve under $\hat{H}_{scat}$.
This evolution is complicated, but for our purposes it turns out to be
sufficient to note that it makes the two-point function decay
with time difference.  In the case where the gas is dilute, we can take
sufficient account of this effect by approximating
\begin{equation}\label{decay}
\hbox{Tr}\Bigl[\hat{\rho}(\beta,\mu)
	\hat{\Psi}_{klm}(t)\hat{\Psi}^\dagger_{k'l'm'}(t')\Bigr]
	\doteq\delta_{kk'}\delta_{ll'}\delta_{mm'}e^{-\gamma_k|t-t'|}\;,
\end{equation}
where $\hat{\rho}(\beta,\mu)$ is the grand canonical ensemble density
operator, and $\gamma_k = \hbar k \sigma d /m$ is the Boltzmann
scattering rate, for
$d=e^{\beta\mu}(m/2\pi\hbar^2\beta)^{3/2}$ the density of the gas\cite{KadanoffBaym}.
The equilibration time for the unbound gas is set by the thermal average of
$\gamma_k$, which we will denote by $\gamma$.

In terms of the time-dependent operators, then, we can write the
interaction picture $\hat{H}_I$ [using the notation $\vec{k} = (k,l,m)$]
\begin{eqnarray}\label{HIs}
\hat{H}_I(t) &=& 2\pi{\hbar^2 a \over m}\Bigl[\sum_{\vec{k}_i}
	[V_3(\vec{k}_i)\hat{\psi}^\dagger_{\vec{k}_1}
	\hat{\psi}_B^\dagger\hat{\psi}_{\vec{k}_2}
	\hat{\psi}_{\vec{k}_3} + \hbox{h.c.}]\nonumber\\
	&& + \sum_{kk'lm}[V_2(k,k',l)\,(
	\hat{\psi}^\dagger_{klm}\hat{\psi}^\dagger_{k'l,-m}
	\hat{\psi}_B\hat{\psi}_B\nonumber\\
&&\qquad\qquad\qquad\qquad + 2\hat{\psi}^\dagger_{klm}\hat{\psi}_{k'lm}
	\hat{n}_B) + \hbox{h.c.}]\nonumber\\
	&& + \sum_{k} V_1(k) \hat{\psi}^\dagger_B(\hat{\psi}^\dagger_B
	\hat{\psi}_{k00} + 
	\hat{\psi}^\dagger_{k00}\hat{\psi}_B)\hat{\psi}_B\Bigr]\;,
\end{eqnarray}
where all the $V_n$ co-efficients are given by integrals of four mode
functions ($u_{klm}$ and $u_B$).

Our goal is to compute the reduced density operator of the bound mode,
from the initial density operator of the bound mode and the reservoir.
In the interaction picture, this is given by
\begin{eqnarray}\label{rhoAt}
\hat{\rho}_B(t) &=& \hbox{Tr}_C\Bigl[ 
	{\cal T}e^{-{i\over\hbar}\int_0^t\!ds\,\hat{H}_I(s)}\nonumber\\
&&\qquad\times
	[\hat{\rho}_B(0)\otimes\hat{\rho}_C(0)]
	\bar{{\cal T}}e^{{i\over\hbar}\int_0^t\!ds\,\hat{H}_I(s)}
	\Bigr]\;,
\end{eqnarray}
where ${\cal T}$ ($\bar{\cal T}$) denote (reverse) time ordering.
We have assumed that the initial density matrix factorizes,
and we will take $\hat{\rho}_C(0)$ to be a grand canonical ensemble
$\hat{\rho}(\beta,\mu)$.
In much the same spirit as that of the influence functional formalism,
we will determine the evolution of the reservoir degrees of freedom and
perform the trace over them before considering the bound mode itself.
The bound mode operators $\hat{\psi}_B(t)$ are not operators in the bath
Hilbert space; but we are apparently prevented from treating them merely
as c-numbers in the bath Hamiltonian, because they are still operators
in the bound mode Hilbert space, and their ordering is significant.
Yet as Feynman observed, this difficulty may be overcome by a
simple notational trick\cite{Feynman}.  

In a slight generalization of Feynman's original ordered
operator calculus, we here add the device of placing primes on all
$\hat{\psi}_B(t)$ operators that appear to the right of the initial
density operators in Eqn.~(\ref{rhoAt}).  We will then evaluate the RHS
of (\ref{rhoAt}) just as if the bound mode operators were not operators
at all, but keep track of their time arguments.  We will afterwards be
able to restore their correct ordering as operators, simply by placing
all unprimed operators time ordered to the left of $\hat{\rho}_B(0)$,
and all primed operators in reverse time order to the right.  This sort
of procedure can always be used, of course; but in general it provides
only an opaque formal solution in terms of time ordered operators.
The convenient features of the present idealized model happen to make
it genuinely powerful here.

Performing the standard equilibrium analysis after momentarily setting
$a\to 0$ confirms that if $\beta E_b >>1$, Bose condensation occurs
at very low fugacity $e^{\beta \mu} \sim e^{-\beta E_b}$.  Assuming
both $\beta$ and $\mu$ to be in this regime, then, we can evaluate
(\ref{rhoAt}) in the dilute gas approximation, where we consider only
independent two particle collisions.  (This approximation affects only the
unbound bath modes: we are not hereby making any assumption concerning
the bound mode particles, or interactions between them!)  We therefore
replace Eqn.~(\ref{rhoAt}) with the dilute gas expression
\begin{eqnarray}\label{rhodg}
\hat{\rho}_B(t) &=& \hat{\rho}_B(0)\exp\Bigl[-{i\over\hbar}
\int_0^t\!ds\hbox{Tr}[\hat{\rho}_C(0)(\hat{H}_I-\hat{H}_I')]\nonumber\\
&& \qquad+{1\over2\hbar^2}\Bigl(
\int_0^t\!ds\hbox{Tr}[\hat{\rho}_C(0)(\hat{H}_I-\hat{H}_I')]\Bigr)^2\\
&-&{1\over\hbar^2}\int_0^t\!\!ds\!\!\int_0^s\!\!ds'\hbox{Tr}
	[\hat{\rho}_C(0)(\hat{H}_I-\hat{H}_I')_s
	(\hat{H}_I-\hat{H}_I')_{s'}]\;,\nonumber
\end{eqnarray}
where the RHS is to be evaluated using the ordering convention described
above for the bound mode operators.  

The second simplification purchased by our idealizations is a time scale
separation, which allows us to collapse the double time integrals in
(\ref{rhodg}) into single integrals, and so obtain Markovian evolution
for the bound mode.  The Boltzman scattering rate is proportional to
$e^{\beta\mu}$, but the condensate grows through collisions between two
gas particles in the trapping well, which (since the well has a finite
volume) occur at a rate proportional to $e^{2\beta\mu}$.  We can therefore
use approximations based on the fact that the condensate evolves much
more slowly than the reservoir equilibrates.  Since there are several
different terms in $\hat{H}_I$, we must in fact employ several variations
of this approximation, and also use the assumption of weak interaction
($\lambda a <<1$).  This analysis will be described in detail elsewhere,
but we will illustrate the least trivial and most important step here.

The following important double time integral from (\ref{rhodg}) represents
a depletion event in which an unbound particle of momentum $k_1$ dislodges
a bound particle, resulting in two unbound particles of momenta $k_2$
and $k_3$:
\begin{eqnarray}\label{term}
&&\int_0^t\!ds\!\int_0^t\!ds'\,e^{-{i\hbar\over2m}(s-s')
	(k_1^2-k_2^2-k_3^2)} e^{-\gamma_{23}|s-s'|}\nonumber\\
&&\qquad\qquad\qquad\qquad\qquad\times 
	\hat{\psi}_B(s)\hat{\psi}^{\dagger'}_B(s')\nonumber\\
&\doteq& \int_0^t\!ds_+\,\hat{\psi}_B(s_+-\epsilon)
	\hat{\psi}_B^{\dagger'}(s_+-\epsilon)\nonumber\\
&&\times
\int_{-s_+}^{s_+}\!ds_-\,e^{-\gamma_{23}|s_-|}
	e^{is_-[{\hbar\over2m}(k_2^2+k_3^2-k_1^2) -{\hat{\Delta}(s_{\scriptscriptstyle +})
	+\hat{\Delta}'(s_+)\over2}]}\nonumber\\
&=& {\cal O}(\gamma) + \pi\int_0^t\!ds\, \hat{\psi}_B(s-\epsilon)\hat{\psi}_B^{\dagger'}(s-\epsilon)\nonumber\\
&&\qquad\times
\delta\bigl({\hbar\over2m}(k_1^2-k_2^2-k_3^2) +{\hat{\Delta}(s)+\hat{\Delta}'(s)\over2}\bigr) \;,
\end{eqnarray}
where $\gamma_{23}$ denotes $\gamma_{k_2}+\gamma_{k_3}$, and in the last
line we have assumed that $t>>\gamma^{-1}$.  (Nothing significant happens
to the bound mode prior to this, anyway.)  The critical step is changing
the time arguments of the operators from $s$ and $s'$ to $s_+=(s+s')/2$.
This can be done because when the exponentials in (\ref{rhodg})
are expanded, the typical time between terms like (\ref{term}) is
proportional to $e^{-2\beta \mu}$, but the difference in each such term
between $s$ and $s'$ can be only of order $e^{-\beta\mu}$.  Our change
of time arguments thus only affects the ordering of $\hat{\psi}_B$ and
$\hat{\psi}^\dagger_B$ operators at order $e^{3\beta\mu}$.  (Ordering
errors involving $\hat{n}_B$ operators must also be considered, but this
can be shown only to renormalize $\gamma_{23}$.)

The delta function that results from the time scale separation
enforces self energy ($H_B + H_{kin}$) conservation to leading
order in $a^2$, and thus realizes the expectation that the dynamics
of a dilute gas is dominated by elastic collisions.  (In terms from
(\ref{rhodg}) with all operators primed or all unprimed, there is
an imaginary part which is not a delta function; but these parts
constitute order $a^2\lambda^2e^{\beta\mu}$ or smaller renormalizations
of the bound mode Hamiltonian, negligible in comparison with the bare
$\hat{H}_B$.)  It is crucially important that the delta function we
obtain has an operator-valued argument.  The legitimacy of this curious
expression, despite its somewhat unnerving appearance, shows the power of
Feynman's ordered operator calculus, for it implies the correct operator
fluctuation-dissipation relation between condensate growth and depletion.
This result seems to be much more difficult to obtain by other methods.

(It is at this point possible to substantially improve our dilute gas
approximation by taking into account depletion of the gas modes during
condensation.  For brevity, however, we will here assume that the number
of condensate particles remains a negligible fraction of the number of
bath particles, so that gas depletion is insignificant.)

Having in similar fashion collapsed all the double integrals in
(\ref{rhodg}), we obtain for the bound mode reduced density operator a
Markovian master equation of Bloch-Lindblad form.  Dropping most of the
$_B$ subscripts, returning to the Schr\"odinger picture where $\hat{\rho}$
is the only time-dependent operator, and abandoning the time-ordered
convention in favour of the ordinary ``as-written'' ordering, it becomes
\begin{eqnarray}\label{master}
{d\hat{\rho}\over dt} &=& {i\over\hbar}[\hat{\rho},\hat{H}_B]
	-\alpha_2\gamma \Bigl(\{\hat{n}^2,\hat{\rho}\}
		-2\hat{n}\hat{\rho}
		\hat{n}\Bigr)\nonumber\\
&&\qquad-\alpha_1{\gamma e^{\beta\mu}\over \beta E_b} \sum_\pm\Bigl(
	\{\hat{Q}_\pm^\dagger\hat{Q}_\pm,\hat{\rho}\}
	-2\hat{Q}_\pm\hat{\rho}\hat{Q}_\pm^\dagger\Bigr)\;,
\end{eqnarray}
where $\alpha_n$ are constants of order unity which can be evaluated
easily to leading order in $(\beta E_b)^{-1}$, and the condensate
feeding and depleting operators $\hat{Q}_\pm$ are defined to be
\begin{eqnarray}\label{Q}
\hat{Q}_+ &\equiv&  \hat{\psi}^\dagger\nonumber\\
\hat{Q}_- &\equiv&  e^{{\beta\over2}(\hbar\hat{\Delta}-\mu)}
\hat{Q}_+^\dagger\;.
\end{eqnarray}
This master equation is our main result; it is correct to second order
in $a\lambda$ and leading order in $e^{\beta\mu}$.  While we have made
free use of special features of our model to derive it, we conclude by
making three observations on the very general physics it represents.

Firstly, we note the terms that do not appear in the master equation.
$\hat{H}_I$ contains vertices (proportional to $V_2$ and $V_3$) which
annihilate one or two bound mode particles and create only one or
two gas particles (or vice versa); but these vertices do not conserve
$H_B+H_{kin}$, and so (apart from some more of the negligible
corrections to $\hat{H}_B$ mentioned above) can only contribute terms of
order $e^{\beta\mu}(a\lambda)^4$.  Excitation of virtual particles from
the condensate may certainly be less negligible in other models; but
an energy gap will always tend to suppress such processes in favour of
elastic ones.

Secondly, it is easy to show that under (\ref{master})
$\hat{\rho}$ relaxes towards the grand canonical ensemble
$Z^{-1}e^{\beta(\mu\hat{n}-\hat{H}_B)}$, with the full, non-linear
$\hat{H}_B$.  We can consider Eqn.~(\ref{Q}) to be an operator FDR which
ensures such relaxation.  Moreover, it is easy to understand this FDR.
As we have noted, the condensate growth process requires two gas particles
to collide over the well, and its strength is thus proportional to
$e^{2\beta\mu}$.  The depletion process requires only one incident gas
particle, and so is of order $e^{\beta\mu}$, but since energy must be
conserved to leading order in $a^2\lambda^2$, the incoming particle must
have high energy in order to dislodge a bound particle.  Depletion is thus
enhanced relative to growth by a (fugacity)$^{-1}$ factor, but suppressed
by a thermal factor.  (Since the processes involve two $\hat{Q}_\pm$
operators, the factors in (\ref{Q}) are the square roots.)  And our
operator delta function in (\ref{term}) implies that the thermal factor
is precisely $e^{\beta\hbar\hat{\Delta}}$, correctly reflecting the fact
that interparticle repulsion lowers the binding energy as $n$ increases.
Furthermore, we can note that $\hbar\hat{\Delta} - \mu = V_{GL}(\hat{n}) -
V_{GL}(\hat{n}-1)$, where
\begin{equation}
V_{GL}(\hat{n}) = \hat{H} - \mu\hat{n} 
		= (-E_b -\mu)\hat\psi^\dagger\hat\psi 
			+ E_r\hat\psi^{\dagger 2}\hat\psi^2
\end{equation}
is the quantized Ginzburg-Landau effective potential.
So the Ginzburg-Landau EP simply describes a competition between capture
of slow gas particles and ejection by fast particles, both processes
proceeding through elastic collisions.

Thirdly, Eqn.~(\ref{master}) also includes the $\alpha_2$ term, which is
due to gas particles scattering off the bound mode condensate via the
elastic part of the $V_2$ vertex in $\hat{H}_I$.  This term represents
diffusion in the phase canonically conjugate to particle number,
as may be seen by expressing (\ref{master}) in the Wigner representation,
and writing $q+ip \sim \psi = \sqrt{n}e^{i\theta}$.  The $\alpha_2$ term
then becomes a second derivative with respect to $\theta$.  While thinking
of the real and imaginary parts of $\psi$ as independent classical
degrees of freedom may suggest that diffusion in $\theta$ should be
of the form ${1\over n}{\partial^2\over\partial\theta^2}$, in fact the
$\alpha_2$ term does not decay with $n$: It does not represent classical
noise, but decoherence.  Yet the argument that a ``Mexican hat'' form for
$V$ immediately implies spontaneous breaking of rotational symmetry in
$\theta$ is actually based on the assumption that, when $n$ is large,
diffusion in $\theta$ becomes negligible.
 
It is certainly possible that decoherence will permit spontaneous
symmetry breaking in other models (though perhaps only for the relative
phase(s) between two or more condensate modes\cite{Naraschewskietal});
but scattering of uncondensed particles is a generic process, which may
typically be expected to produce phase diffusion terms resembling that
obtained here.  These terms may well fail in some way to prevent symmetry
breaking in other models, but they will not obviously vanish due to
mere kinematics as the number of condensed particles becomes large.
This strongly suggests the following general conclusion.

Spontaneous symmetry breaking, in which states of definite particle number
are replaced by states where the phase which is conjugate to particle
number is definite instead, is important in many areas of physics.
A ``Mexican hat'' form of effective potential may be a necessary condition
for such SSB, but it is not sufficient, because diffusion associated
with decoherence occurs in addition to the diffusion characterized by
the effective potential.  This implies that emergence of the symmetry
breaking order parameter in these cases is not a quasi-classical process,
but is an especially non-trivial instance of classicality emerging from
quantum mechanics.

\begin{center}
{\bf Acknowledgements}
\end{center}

The author appreciated the hospitality of the Institute for Theoretical
Physics at UCSB, supported by NSF grant PHY94-07194, during the later
stages of this work.  He thanks A.~Gill, E.~Mottola, A.~Imamoglu,
P.~Zoller and S.~Habib for valuable discussions.

\end{document}